\documentclass[english,aps,pre, reprint, twocolumn]{revtex4}
\usepackage[T1]{fontenc}
\usepackage[latin9]{inputenc}
\usepackage{amsmath}
\usepackage{amssymb}
\usepackage{graphicx}

\makeatletter
\@ifundefined{textcolor}{}
{%
 \definecolor{BLACK}{gray}{0}
 \definecolor{WHITE}{gray}{1}
 \definecolor{RED}{rgb}{1,0,0}
 \definecolor{GREEN}{rgb}{0,1,0}
 \definecolor{BLUE}{rgb}{0,0,1}
 \definecolor{CYAN}{cmyk}{1,0,0,0}
 \definecolor{MAGENTA}{cmyk}{0,1,0,0}
 \definecolor{YELLOW}{cmyk}{0,0,1,0}
}

\makeatother

\usepackage{babel}
\begin{document}

\preprint{\%This line only printed with preprint option}

\title{A microscopic approach to understand Brownian dynamics in viscoelastic
fluid}

\author{Shuvojit Paul}

\affiliation{Indian Institute of Science Education and Research, Kolkata }

\author{Basudev Roy}
\email{basudev@gmail.com}

\affiliation{Indian Institute of Technology Madras}

\author{Ayan Banerjee}
\email{ayan@iiserkol.ac.in}

\affiliation{Indian Institute of Science Education and Research, Kolkata }
\begin{abstract}
We present an entirely microscopic formulation of viscoleasticity
of a fluid starting from the microscopic Stokes-Oldroyd B Model assuming
instantaneous hydrodynamic friction, and show that linearization leads
to a form for the frequency dependent viscosity that can be directly
applied to the Langevin equation. Interestingly, the calculated expression
of viscosity can be directly mapped to the Jeffreys' model which is
essentially macroscopic in nature with the bulk viscoelasticity of
the fluid being characterized by a complex elastic modulus $G(\omega)$.
Further, we demonstrate that the concerned Green's function is same
as that in an incompressible, low Reynold's number Newtonian fluid
with the simple incorporation of frequency dependence in the viscosity
term. We proceed to evaluate the trajectory of a free Brownian particle
in a viscoelastic environment using our formalism, and calculate parameters
such as the power spectral density, the autocorrelation function and
the mean-square displacement, which we then extend to the particle
confined in a harmonic potential in the fluid.
\end{abstract}
\maketitle

\section{introduction}

The study of complex fluids is of paramount importance primarily since
biological entities which sustain life exist in such fluidic environments
- whose complexity is typically manifested by a `viscoelastic' nature
\cite{brust2013rheology,ayala2016rheological,prado2015viscoelastic,larson1999structure,ferry1980viscoelastic},
wherein the fluid exhibits both viscous and elastic properties \cite{doi1988theory}.
Understandably, there exists a strong interest in the scientific community
to understand the exact nature of viscoelastic fluids and to measure
their rheological properties. Most models that attempt to analyze
visco-elasticity are essentially ``macroscopic'' in nature, and
attempt to understand the fluid in terms of its bulk viscous and elastic
properties, and their response to external stresses. The foremost
and simplest among the models developed is the Maxwell model \cite{gotze2008complex,boon1980molecular},
which has been improved further into the generalized Maxwell or Jeffreys'
model \cite{grimm2011brownian}, and can provide the relation between
stress and shear-rate in linear viscoelastic fluids, which can then
be used to calculate frequency dependent viscoelastic parameters of
the medium. These measurements facilitate ``macrorheological'' measurements
of fluid parameters using commercial rheometers that require several
ml of sample volume, and typically determine the rheological properties
at low frequencies \cite{derkach2009rheology}. Thus, the rheological
properties of viscoelastic fluids are expressed in terms of a frequency
dependent complex elastic modulus $G(\omega)$, whose real part represents
storage and imaginary part represents loss. Finally, $G(\omega)$
gives rise to a complex frequency dependent viscosity $\eta(\omega)$
using $\eta(\omega)=\dfrac{G(\omega)}{-i\omega}$ - the real and imaginary
components of which have similar associations as those of $G(\omega)$.

In contrast, a 'microscopic' approach towards comprehending the viscoelastic
properties of a complex fluid would attempt to link the basic constituents
of the fluid \cite{le2012micro}- often conceptualized as polymer
chains of different length scales embedded and moving inside a Newtonian
viscous fluid - and the macroscopic rheological properties of the
fluid. Such a link would enable the synthesis of complex fluids much
more simpler, since although the microscopic features of a polymer
chain are often well known, the exact effect they would have in determining
the macroscopic rheological properties of the fluid is very difficult
to predict. Microscopic modeling would thus be an excellent predictive
tool, which remains not very accessible due to the associated high
computational cost involved with the introduction of additional microscopic
variables. The rheological effects of the polymer chains can, however,
be studied on a microscopic scale, by determining the the trajectory
of a Brownian particle embedded in the fluid, which is manifested
in the mean squared displacement (MSD) of the particle. The viscous
and elastic properties may then be illustrated by considering the
particle to be attached to a dashpot and a spring, respectively. In
this approach, the Maxwell model provides a comprehensive description
of the viscoelastic effects of the fluid on the particle trajectory
assuming instantaneous hydrodynamic friction. A so-called 'Maxwell'
fluid is also characterized by a characteristic time-scale $\tau_{m}$,
which marks a transition from a high frequency elastic regime to a
low-frequency purely viscous regime. The effects of the solvent's
dissipation is incorporated in the generalized Maxwell's or Jeffrey's
model which interprets this as a background viscosity $\eta_{\infty}$,
and also redefines the frequency regimes where purely viscous (very
low and very high frequencies) and elastic (intermediate frequencies)
behaviour is displayed by the fluid. Then, the drag force on the particle
is evaluated by solving the Navier Stokes equations with the no-slip
boundary condition and integrating over the stress tensor on the particle
surface.This treatment enables ``microrheology'', which is growing
very rapidly due to its capability to explore small-scale intrinsic
properties of complex fluids over a wide frequency band \cite{mason1997particle,mason1995optical,dasgupta2002microrheology,tassieri2010measuring,wilking2008optically,gomez2014probing,rich2011nonlinear}.
Microrheology - as the name suggests - employs Brownian probes embedded
in the medium and measures their position autocorrelations induced
by thermal fluctuations (passive microrheology \cite{mason1995optical,mason1997particle,mason2000estimating}),
or response to external forces (active microrheology \cite{guzman2008situ,cicuta2007microrheology,kimura2009microrheology}),
in order to determine the viscoelastic properties of the fluid. Now,
an interesting exercise would be to consider viscoelasticity at the
level of the polymer chains themselves, as well as incorporate solvent
properties such as incompressibility into a model that would express
the effects of viscoelasticity into a simple expression of frequency
dependent viscosity, and compare it with the existing ``macroscopic''
models, as well as determine the trajectory of Brownian probes, both
free and in confinement enabled by optical traps - which would thus
cast the problem in a microrheological context.

In this paper, we perform this exercise rather thoroughly. We start
with the Stokes-Oldroyd-B model \cite{thomases2007emergence,thomases2011stokesian}
- an existing microscopic model of viscoelasticity - and linearize
it to understand it's behavior to small external perturbations. We
then formulate a hydrodynamic Green's function - which, finally yields
a frequency dependent complex viscosity. Interestingly, though we
linearize the quasi-linear Stokes-Oldroyd-B equation, the expression
for the complex viscosity differs from that obtained from the stress-strain
relation of Maxwell model. However, we observe a qualitative similarity
in terms of the viscosity contributions of the solvent and the polymer
chains in the total frequency dependent expression of viscosity for
the fluid. Furthermore, we determine the thermal trajectory of a Brownian
particle - both free and confined in a harmonic oscillator potential
as is the case in optical tweezers - in a low-Reynold's number, incompressible,
viscoelastic fluid. It is clear that the statistical properties of
the phenomena are connected to the rheological properties of the fluid
and thus facilitate accurate microrheology.

\section{Theory}

The Stokes-Oldroyd-B equations - having a microscopic origin - for
an incompressible, low Reynold's number fluid are given by
\begin{gather}
\boldsymbol{\nabla}\cdot\boldsymbol{u}=0\label{eq:1}\\
-\boldsymbol{\nabla}p+\mu_{s}\Delta\boldsymbol{u}=-\boldsymbol{\nabla}\cdot\mathbb{S}-\boldsymbol{f}\label{eq:2}\\
\frac{\partial\mathbb{S}}{\partial t}+\left(\boldsymbol{u}\cdot\boldsymbol{\nabla}\right)\mathbb{S}-\left(\mathbb{S}\boldsymbol{\nabla}\boldsymbol{u}+(\boldsymbol{\nabla}\boldsymbol{u})^{T}\mathbb{S}\right)+\frac{\mathbb{S}}{\lambda}=\nonumber \\
\frac{\mu_{p}}{\lambda}\left(\boldsymbol{\nabla}\boldsymbol{u}+(\boldsymbol{\nabla}\boldsymbol{u})^{T}\right)\label{eq:3}
\end{gather}

where, $\boldsymbol{u}$, $p$, $\mathbb{S}$, $\boldsymbol{f}$ are
the fluid velocity, pressure, polymer contribution to the stress tensor,
and external force, respectively. $\mu_{s}$ is the zero-frequency
solvent viscosity, $\mu_{p}$ is the zero-frequency polymer viscosity
and $\lambda$ is the polymer relaxation time scale. $\mu_{p}=\mu_{o}-\mu_{s}$
where $\mu_{0}$ is the viscosity of the solution at $\omega=0$.
We linearize above equations by considering $\boldsymbol{f}=\epsilon\boldsymbol{g}$,
$\boldsymbol{u}=\epsilon\boldsymbol{u}'$, $p=\epsilon p'$ and $\mathbb{S}=\epsilon\mathbb{T}$
where $\epsilon\ll1$. Now \eqref{eq:2} and \eqref{eq:3} become

\begin{gather}
-\boldsymbol{\nabla}p'+\mu_{s}\Delta\boldsymbol{u}'+\boldsymbol{g}=-\boldsymbol{\nabla}\cdot\mathbb{T}\label{eq:4}\\
\frac{\partial\mathbb{T}}{\partial t}-\frac{\mu_{p}}{\lambda}\left(\boldsymbol{\nabla}\boldsymbol{u}+(\boldsymbol{\nabla}\boldsymbol{u})^{T}\right)+\frac{\mathbb{T}}{\lambda}=0\label{eq:5}
\end{gather}

Clearly, \eqref{eq:5} relates stress tensor with shear-rate tensor
as it is in Maxwell model of viscoelastic fluid. Equations \eqref{eq:1},
\eqref{eq:4} and \eqref{eq:5} can be written in component form as

\begin{gather}
\partial_{i}u'_{i}=0\label{eq:6}\\
-\partial_{i}p'+\mu_{s}\left(\partial_{i}^{2}u_{i}'+\partial_{j}^{2}u_{i}'+\partial_{k}^{2}u_{i}'\right)+g_{i}=-\partial_{j}\mathbb{T}_{ij}\label{eq:7}\\
\frac{\partial\mathbb{T}_{ij}}{\partial t}-\frac{\mu_{p}}{\lambda}\left(\partial_{i}u_{j}'+\partial_{j}u_{i}'\right)+\frac{\mathbb{T}_{ij}}{\lambda}=0\label{eq:8}
\end{gather}

where Einsten summation has been used. Now, after performing a Fourier
transform on \eqref{eq:6}, \eqref{eq:7} and \eqref{eq:8}, we obtain

\begin{gather}
k_{i}\widetilde{u_{i}}'=0\label{eq:9}\\
-ik_{i}\widetilde{p}'-\mu_{s}k^{2}\widetilde{u_{i}}'+\widetilde{g_{i}}=-ik_{j}\widetilde{\mathbb{T}}_{ij}\label{eq:10}\\
-i\omega\mathbb{\widetilde{T}}_{ij}-\frac{\mu_{p}}{\lambda}\left(ik_{i}\widetilde{u}_{j}'+ik_{j}\widetilde{u}_{i}'\right)+\frac{\mathbb{\widetilde{T}}_{ij}}{\lambda}=0\label{eq:11}
\end{gather}

where the rules of four dimensional Fourier transforms were used.
Multiplying $k_{j}$ with \eqref{eq:11} we obtain,

\[
-i\omega k_{j}\mathbb{\widetilde{T}}_{ij}-\frac{\mu_{p}}{\lambda}\left(ik_{i}k_{j}\widetilde{u}_{j}'+ik_{j}k_{j}\widetilde{u}_{i}'\right)+\frac{k_{j}\mathbb{\widetilde{T}}_{ij}}{\lambda}=0
\]

Now, using \eqref{eq:9} we finally obtain, 
\begin{equation}
k_{j}\mathbb{\widetilde{T}}_{ij}=i\frac{\mu_{p}}{\lambda}\frac{k^{2}}{\left(-i\omega+\frac{1}{\lambda}\right)}\widetilde{u}_{i}'\label{eq:12}
\end{equation}

so that, after plugging \eqref{eq:12} into \eqref{eq:10} we have,

\begin{eqnarray}
-ik_{i}\widetilde{p}'-\mu_{s}k^{2}\widetilde{u_{i}}'+\widetilde{g_{i}} & = & \left(\frac{\mu_{p}}{\lambda}\frac{k^{2}}{\left(-i\omega+\frac{1}{\lambda}\right)}\widetilde{u}_{i}'\right)\label{eq:13}
\end{eqnarray}

By performing an inverse Fourier transform of \eqref{eq:13} to come
back to real space and taking care of all other components, we have 

\begin{equation}
-\boldsymbol{\nabla}p'(\boldsymbol{r},\omega)+\biggl[\mu_{s}+\frac{\mu_{p}}{\left(-i\omega\lambda+1\right)}\biggr]\Delta\boldsymbol{u}'(\boldsymbol{r},\omega)+\boldsymbol{g}(\boldsymbol{r},\omega)=\boldsymbol{0}\label{eq:14}
\end{equation}

We can also perform an inverse Fourier transform to \eqref{eq:9}
to get 

\begin{equation}
\boldsymbol{\nabla}\cdot\boldsymbol{u}'(\boldsymbol{r},\omega)=0\label{eq:15}
\end{equation}

\eqref{eq:14} and \eqref{eq:15} are similar to the stokes equation
in low Reynold's number, incompressible Newtonian fluid which are
given below.

\begin{gather}
-\boldsymbol{\nabla}P+\mu\Delta\boldsymbol{v}+\boldsymbol{F}=\boldsymbol{0}\label{eq:16}\\
\boldsymbol{\nabla}\cdot\boldsymbol{v}=0\label{eq:17}
\end{gather}

$P,\,\boldsymbol{v,}\,\boldsymbol{F}$ are pressure, velocity, external
force, respectively, while $\mu$ is the fluid viscosity. The above
two equations can be written in $(\boldsymbol{r},\omega)$ space where
they appear as

\begin{gather}
-\boldsymbol{\nabla}P(\boldsymbol{r},\omega)+\mu\Delta\boldsymbol{v}(\boldsymbol{r},\omega)+\boldsymbol{F}(\boldsymbol{r},\omega)=\boldsymbol{0}\label{eq:18}\\
\boldsymbol{\nabla}\cdot\boldsymbol{v}(\boldsymbol{r},\omega)=0\label{eq:19}
\end{gather}

\begin{figure*}[!t]
\includegraphics[scale=0.5]{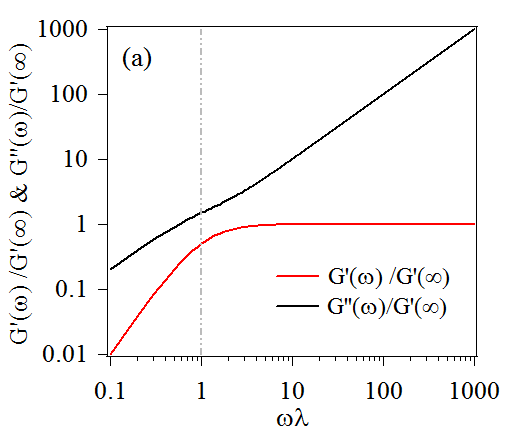}\includegraphics[scale=0.5]{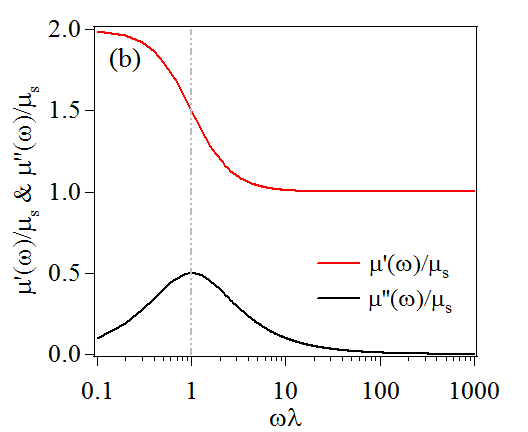}\caption{The real and imaginary part of the dynamic complex modulus $G'$ and
$G''$ are plotted against the angular frequency $\omega$ in (a),
while the frequency dependency of $\mu(\omega)$ is shown in (b).
$\mu_{p}$ and $\mu_{s}$ are chosen to be equal so that the polymer
concentration in the solution is relatively low so that the viscoelasticity
of the fluid is manifested clearly, but is not large enough for the
fluid to display non-linear characteristics. The dotted gray line
represents $1/\lambda$ frequency. Clear, this is the transition frequency
of the fluid where viscous nature starts to dominate over the elastic
nature. }
\label{fig:1}

\end{figure*}

Hence, we can compare \eqref{eq:14} and \eqref{eq:18}, and can conclude
that

\begin{equation}
\mu(\omega)=\biggl[\mu_{s}+\frac{\mu_{p}}{\left(-i\omega\lambda+1\right)}\biggr]\label{eq:20}
\end{equation}

is the expression of frequency dependent viscosity in an incompressible,
low Reynold's number viscoelastic fluid. Thus, the frequency dependent
dynamic complex modulus is $G^{*}(\omega)=-i\omega\mu(\omega)=G'(\omega)-iG''(\omega).$
The real part $G'(\omega)$ represents the storage modulus and $G''(\omega)$
refers to the loss modulus of the fluid. The result can be compared
to the expression obtained from the Maxwell model \cite{grimm2011brownian},
where the frequency dependent viscosity is given by
\[
\mu_{M}(\omega)=\frac{\eta_{0}}{1-i\omega\tau_{M}}
\]

where $\eta_{0}$ is the zero-frequency viscosity, and $\tau_{M}$
is the Maxwell time described earlier. This is modified to the generalized
Maxwell model or Jeffrey's model which gives 
\[
\mu_{M}(\omega)=\eta_{\infty}+\frac{\eta_{0}}{1-i\omega\tau_{M}}
\]

where $\eta_{\infty}$ is the viscosity of the solution at $\omega\rightarrow\infty$.
It is clear that the inclusion of $\eta_{\infty}$ is performed in
order to ensure that a physically meaningful result is obtained at
very large values of $\omega$, where the behaviour of the fluid is
essentially viscous. However, $\eta_{\infty}$ is defined as the background
viscosity that arises due to the constituent mesoscopic particles
of the viscoelastic material being immersed in a solvent \cite{grimm2011brownian}.
This can directly be mapped to the solvent viscosity $\mu_{s}$ that
we define in our approach. Thus, it is interesting to note that while
the Maxwell model and Jeffrey's model both originate macroscopically,
and do not directly include the incompressibility (Eq.\eqref{eq:1})
of the fluid and the momentum balance condition (Eq.\eqref{eq:2}),
our microscopic model which considers both these conditions, give
rise to very similar results. Note that $\lambda$ in our model is
the ratio of viscosity of the solution to its elasticity, so that
it has a dimension of time, and can again be interpreted to be of
similar significance to the Maxwell model, since when the medium becomes
purely viscous then $\lambda\rightarrow\infty$ and from \eqref{eq:20},
$\mu(\omega)=\mu_{s}$. This is identical to the behaviour of $\tau_{M}$
in the Maxwell model(s). $G^{*}(\omega)$ and $\mu(\omega)$ are plotted
in Fig.\ref{fig:1}. It is clear from the figure that the fluid becomes
purely viscous as $\omega\rightarrow\infty$ as the real part of $\mu$
approaches $\mu_{s}$ and the imaginary part tends to zero. 

\subsection{Green's function calculation:}

Taking inner product of equation \eqref{eq:13} with $ik_{i}$ and
using equation \eqref{eq:9} we obtain,

\[
k^{2}\widetilde{p}'_{i}+ik_{i}\widetilde{g}_{i}=0
\]

In three dimensions, the above equation can be written as

\begin{eqnarray}
k^{2}\widetilde{p}' & = & -i\boldsymbol{k}\cdot\widetilde{\boldsymbol{g}}\nonumber \\
\widetilde{p}'(\boldsymbol{k},\omega) & = & -\frac{i\boldsymbol{k}\cdot\widetilde{\boldsymbol{g}}}{k^{2}}.\label{eq:21}
\end{eqnarray}

Using Eq.\eqref{eq:21} into the three dimensional form of Eq.\eqref{eq:13},
we get

\begin{gather}
\boldsymbol{k}\left(\frac{\boldsymbol{k}\cdot\widetilde{g}}{k^{2}}\right)+\mu(\omega)k^{2}\widetilde{\boldsymbol{u}}'=\widetilde{g}\nonumber \\
\widetilde{\boldsymbol{u}}'(\boldsymbol{k},\omega)=\frac{1}{\mu(\omega)k^{2}}\biggl[\widetilde{g}-\boldsymbol{k}\left(\frac{\boldsymbol{k}\cdot\widetilde{g}}{k^{2}}\right)\biggr]\label{eq:22}
\end{gather}

So,
\begin{gather}
p'(\boldsymbol{r},\omega)=\frac{1}{(2\pi)^{3}}\int_{\mathbb{R}^{3}}d\boldsymbol{k}\left(-\frac{i\boldsymbol{k}\cdot\widetilde{\boldsymbol{g}}}{k^{2}}\right)\exp\left(i\boldsymbol{k}\cdot\boldsymbol{r}\right)\label{eq:23}\\
\boldsymbol{u}'(\boldsymbol{r},\omega)=\frac{1}{(2\pi)^{3}\mu(\omega)}\int_{\mathbb{R}^{3}}d\boldsymbol{k}\frac{1}{k^{2}}\biggl[\widetilde{g}-\boldsymbol{k}\left(\frac{\boldsymbol{k}\cdot\widetilde{\boldsymbol{g}}}{k^{2}}\right)\biggr]\exp\left(i\boldsymbol{k}\cdot\boldsymbol{r}\right)\label{eq:24}
\end{gather}

If $\boldsymbol{g}$ is a constant force then Eq.\eqref{eq:23} gets
solved to

\begin{equation}
p'(\boldsymbol{r},\omega)=-\boldsymbol{g}\cdot\nabla\left(\frac{1}{4\pi r}\right)=\frac{\boldsymbol{r}\cdot\boldsymbol{g}}{4\pi r^{3}}\label{eq:25}
\end{equation}

where we used 
\[
\nabla\left(\frac{1}{4\pi r}\right)=\frac{i}{(2\pi)^{3}}\int_{\mathbb{\mathbb{R}}^{3}}d\boldsymbol{k}\frac{\boldsymbol{k}}{k^{2}}\exp\left(i\boldsymbol{k}\cdot\boldsymbol{r}\right)
\]

and Eq.\eqref{eq:24} becomes 
\begin{equation}
\boldsymbol{u}'(\boldsymbol{r},\omega)=\frac{\boldsymbol{g}}{4\pi\mu(\omega)r}-\frac{\boldsymbol{g}}{\mu(\omega)}\cdot\boldsymbol{\nabla}\boldsymbol{\nabla}\left(\frac{r}{8\pi}\right)\label{eq:26}
\end{equation}

where we make use of
\[
\boldsymbol{\nabla}\boldsymbol{\nabla}\left(\frac{r}{8\pi}\right)=\frac{1}{(2\pi)^{3}}\int_{\mathbb{R}^{3}}d\boldsymbol{k}\frac{\boldsymbol{k}\boldsymbol{k}}{k^{4}}\exp\left(i\boldsymbol{k}\cdot\boldsymbol{r}\right)
\]

and $\boldsymbol{k}\boldsymbol{k}\cdot\boldsymbol{g}=(\boldsymbol{g}\cdot\boldsymbol{k})\boldsymbol{k}$.
Since, $\boldsymbol{\nabla}r=\frac{\boldsymbol{r}}{r}$ and the tensor
product $\boldsymbol{\nabla}\boldsymbol{r=}\mathbb{I}$, we have

\begin{equation}
\boldsymbol{u}'(\boldsymbol{r},\omega)=\frac{1}{8\pi\mu(\omega)r}\left(\mathbb{I}+\frac{\boldsymbol{rr}}{r^{2}}\right)\cdot\boldsymbol{g}\label{eq:27}
\end{equation}

Now, in general, the velocity field at a point can be expressed as
\[
\boldsymbol{u}'(\boldsymbol{r},\omega)=\int\mathbb{G}(\boldsymbol{r}-\boldsymbol{r}',\omega)\cdot\boldsymbol{g}(\boldsymbol{r}',\omega)d\boldsymbol{r}'
\]

where
\begin{equation}
\mathbb{G}(\boldsymbol{r})=\frac{1}{8\pi\mu(\omega)r}\left(\mathbb{I}+\frac{\boldsymbol{rr}}{r^{2}}\right)\label{eq:28}
\end{equation}

is the hydrodynamic Green's function or Oseen tensor in a viscoelastic
fluid. $\boldsymbol{g}(\boldsymbol{r}',\omega)$ is a force acting
only in a single point $\boldsymbol{r}'$on the fluid. Under no-slip
boundary condition, the above Green's function can be used to calculate
the hydrodynamic drag force on a spherical particle \cite{kim2013microhydrodynamics}.
Since, $\mu(\omega)$ is only a function of frequency and has no spatial
dependency, the translational drag force will be
\[
\boldsymbol{F}=6\pi\mu(\omega)a_{0}\bigg[\left(1+\frac{a_{0}^{2}}{6}\nabla^{2}\right)\boldsymbol{u}_{1}-(\boldsymbol{U}-\boldsymbol{u}^{\infty})\bigg]
\]

where $a_{0}$ is the radius of the particle, $\boldsymbol{u}_{1}$
is the velocity field caused by other means evaluated at the sphere
center, $\boldsymbol{U}$ is the velocity of the particle and $\boldsymbol{u}^{\infty}$
is the velocity of the uniform background flow evaluated at the center
of the sphere. 

\subsection{Brownian motion in a harmonic oscillator potential:}

The generalized Langevin equation describing the one dimensional trajectory
of a Brownian spherical particle of mass $m$ in a harmonic oscillator
potential in this fluid is given by

\begin{equation}
m\ddot{x}(t)=-\int_{-\infty}^{t}\gamma(t-t')\dot{x}(t')dt'-kx(t)+\xi(t)\label{eq:29}
\end{equation}

where, $\gamma(t)$ is the time dependent friction coefficient, $k$
is the trap stiffness and $\xi(t)$ is the correlated thermal noise
with correlation $\left<\xi(t)\xi(t')\right>=2k_{B}T\gamma(t-t')$.
Since, the fluid has low Reynold's number and the mass of the colloid
is very small, momentum of the particle relaxes in a negligible time
scale so that the effect of inertia is negligible, and the above equation
in frequency domain can be written as:
\begin{eqnarray}
0 & = & i\omega\gamma(\omega)x(\omega)-kx(\omega)+\xi(\omega)\\
x(\omega) & = & \frac{\xi(\omega)}{\left(-i\omega\gamma(\omega)+k\right)}
\end{eqnarray}

Here, $\gamma(\omega)=6\pi\mu(\omega)a_{0}$. So, the power spectral
density (PSD) is given by,

\begin{equation}
\left<x(\omega)x^{*}(\omega)\right>=\frac{2k_{B}T}{\gamma_{0}}\frac{\left(\frac{\left(1+\frac{\mu_{p}}{\mu_{s}}\right)}{\lambda^{2}}+\omega^{2}\right)}{\biggl[\left(\frac{k}{\gamma_{0}\lambda}-\omega^{2}\right)^{2}+\omega^{2}\left(\frac{k}{\gamma_{0}}+\frac{1}{\lambda}\left(1+\frac{\mu_{p}}{\mu_{s}}\right)\right)^{2}\biggr]}\label{eq:32}
\end{equation}

where, $\gamma_{0}=6\pi\mu_{s}a_{0}$ and we used $\left<\xi(\omega)\xi^{*}(\omega)\right>=2k_{B}T\times Re[\gamma(\omega)]$
The position autocorrelation function (ACF) is the inverse Fourier
transform of the PSD, and is given by

\begin{gather}
\left<x(\tau)x(0)\right>=\frac{1}{\nu}\Biggl[\frac{a-\frac{b}{4}(c-\nu)^{2}}{\biggl\{\left(\frac{c-\nu}{2}\right)^{2}+\left(\frac{c-\nu}{2}\right)c+\omega_{0}\biggr\}}\exp\biggl(-\left(\frac{c-\nu}{2}\right)\tau\biggr)\nonumber \\
+\frac{\frac{b}{4}(c+\nu)^{2}-a}{\biggl\{\left(\frac{c+\nu}{2}\right)^{2}+\left(\frac{c+\nu}{2}\right)c+\omega_{0}\biggr\}}\exp\biggl(-\left(\frac{c+\nu}{2}\right)\tau\biggr)\Biggr]\label{eq:33}
\end{gather}

where, $a=\frac{2k_{B}T}{\gamma_{0}\lambda^{2}}\left(1+\frac{\mu_{p}}{\mu_{s}}\right)$,
$b=\frac{2k_{B}T}{\gamma_{0}}$, $\omega_{0}=\frac{k}{\gamma_{0}\lambda}$,
$c=\frac{k}{\gamma_{0}}+\frac{1}{\lambda}\left(1+\frac{\mu_{p}}{\mu_{s}}\right)$
and $\nu=\sqrt{c^{2}-4\omega_{0}}$. The mean-square displacement
(MSD) is related to the ACF as
\begin{gather}
\left\langle \Delta x^{2}(\tau)\right\rangle =2\left[\left\langle x^{2}(0)\right\rangle -\left\langle x(\tau)x(0)\right\rangle \right]\nonumber \\
=\frac{2}{\nu}\Biggl[\frac{a-\frac{b}{4}(c-\nu)^{2}}{\biggl\{\left(\frac{c-\nu}{2}\right)^{2}+\left(\frac{c-\nu}{2}\right)c+\omega_{0}\biggr\}}\Biggl(1-\exp\biggl(-\left(\frac{c-\nu}{2}\right)\tau\biggr)\Biggr)\nonumber \\
+\frac{\frac{b}{4}(c+\nu)^{2}-a}{\biggl\{\left(\frac{c+\nu}{2}\right)^{2}+\left(\frac{c+\nu}{2}\right)c+\omega_{0}\biggr\}}\Biggl(1-\exp\biggl(-\left(\frac{c+\nu}{2}\right)\tau\biggr)\Biggr)\Biggr]\label{eq:34}
\end{gather}

\begin{figure*}[!t]
\includegraphics[scale=0.4]{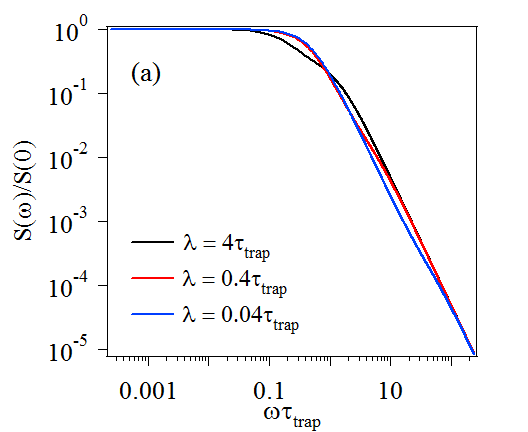}\includegraphics[scale=0.4]{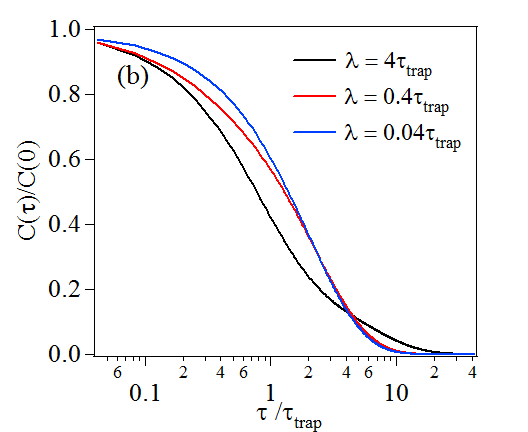}\includegraphics[scale=0.4]{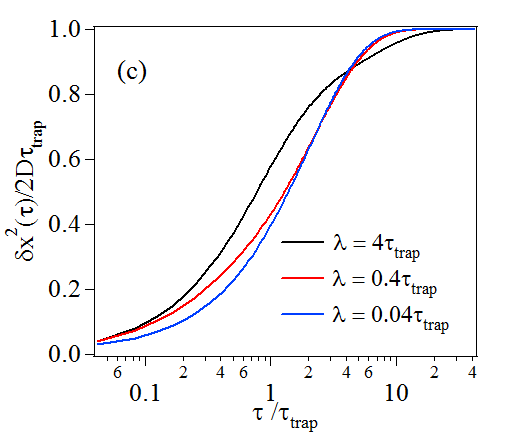}

\caption{PSD, ACF and MSD are plotted with respect to angular frequency and
time-lag in subfigures (a), (b) and (c) respectively for different
time constants $\lambda$. The trap time constant is given by $\tau_{trap}=\gamma_{0}/k$. }

\label{fig:2}
\end{figure*}

$\lambda$ represents the crossover time of the fluid from viscous
to elastic domain. It is clear that if $\lambda\rightarrow\infty$
then 
\begin{gather}
\left<x(\omega)x^{*}(\omega)\right>=\frac{2k_{B}T}{\gamma_{0}}\frac{1}{\omega^{2}+\left(\frac{k}{\gamma_{0}}\right)^{2}}\label{eq:35}\\
\left<x(\tau)x(0)\right>=\frac{k_{B}T}{k}\exp\left(-\frac{k}{\gamma_{0}}\tau\right)\label{eq:36}\\
\left\langle \Delta x^{2}(\tau)\right\rangle =\frac{2k_{B}T}{k}\Biggl(1-\exp\left(-\frac{k}{\gamma_{0}}\tau\right)\Biggr)\label{eq:37}
\end{gather}

which are the expressions in a viscous medium in an optical tweezers
of stiffness $k$. The power spectral density, auto correlation function
and the mean-square displacement functions are plotted for different
$\lambda$ in Fig.\ref{fig:2}.

\subsubsection{Free Brownian particle:}

The above expressions for free Brownian particles in such fluid can
be obtained making $k\rightarrow0$. Then clearly, the above expressions
of PSD, ACF and MSD become

\begin{gather}
\left<x(\omega)x^{*}(\omega)\right>=\frac{2k_{B}T}{\gamma_{0}}\frac{\left(\frac{\left(1+\frac{\mu_{p}}{\mu_{s}}\right)}{\lambda^{2}}+\omega^{2}\right)}{\omega^{2}\biggl[\omega^{2}+\left(\frac{\left(1+\frac{\mu_{p}}{\mu_{s}}\right)}{\lambda}\right)^{2}\biggr]}\label{eq:38}\\
\left<x(\tau)x(0)\right>=\frac{k_{B}T}{\gamma_{0}\pi}\Biggl[\frac{\pi\lambda}{\left(1+\frac{\mu_{p}}{\mu_{s}}\right)^{2}}\exp\left(-\frac{\left(1+\frac{\mu_{p}}{\mu_{s}}\right)}{\lambda}\tau\right)\nonumber \\
-\frac{\pi}{\left(1+\frac{\mu_{p}}{\mu_{s}}\right)}\tau\Biggr]\label{eq:39}\\
\left\langle \Delta x^{2}(\tau)\right\rangle =\frac{2k_{B}T}{\gamma_{0}\pi}\Biggl[\frac{\pi\lambda}{\left(1+\frac{\mu_{p}}{\mu_{s}}\right)^{2}}\nonumber \\
-\frac{\pi\lambda}{\left(1+\frac{\mu_{p}}{\mu_{s}}\right)^{2}}\exp\left(-\frac{\left(1+\frac{\mu_{p}}{\mu_{s}}\right)}{\lambda}\tau\right)+\frac{\pi}{\left(1+\frac{\mu_{p}}{\mu_{s}}\right)}\tau\Biggr]\label{eq:40}
\end{gather}

From equation Eq.\eqref{eq:40}, it is clear that $\left\langle \Delta x^{2}(\tau)\right\rangle =2D\tau$
when $\lambda\rightarrow\infty$ and $\mu_{p}=0$ which is the case
for a free moving particle in a viscous medium where $D=\frac{k_{B}T}{\gamma_{0}}$
is the diffusion coefficient. 

\subsubsection{Implications in rheology:}

\begin{figure}[!bph]
\includegraphics[scale=0.45]{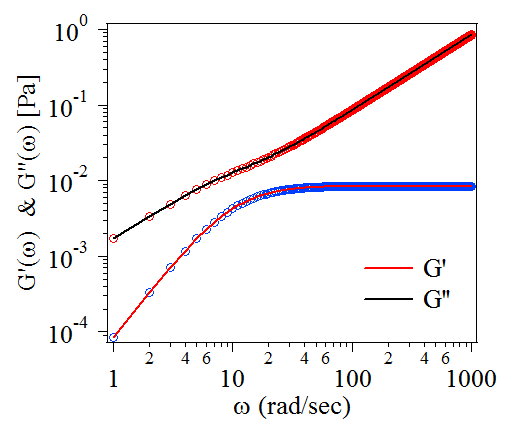}

\caption{Solid lines represent $G^{*}(\omega)$ by Eq.\eqref{eq:32} and the
open circles represent that obtained using Eq.\eqref{eq:41}. $k=1$
$\mu N/m$ was chosen which is relevant for optical tweezers. Other
parameters were taken so that experimental conditions fit well. $\mu_{s}=\mu_{p}=0.00085$
Pa.s, $\lambda=0.1\,s$ were chosen. }

\label{fig:3}
\end{figure}

Assuming that the bulk Laplace-frequency dependent viscosity of the
fluid $\widetilde{\eta}(s)$ is linearly proportional to the memory
function $\widetilde{\xi}(s)$, the complex modulus can be represented
as 
\begin{equation}
G^{*}(\omega)=\frac{k}{6\pi a_{0}}\Bigg[\frac{2\left<x^{2}\right>}{i\omega\left<\Delta\hat{x^{2}}(\omega)\right>}-1\Bigg]\label{eq:41}
\end{equation}

where, $\left<x^{2}\right>$ is the variance and $\left<\Delta\hat{x^{2}}(\omega)\right>$
is the Fourier transform of the mean-square displacement $\left\langle \Delta x^{2}(\tau)\right\rangle $.
In practice, this Fourier transform - given there exist only a finite
set of data points over a finite time domain - is non trivial \cite{tassieri2010measuring}.
Several methods have thus been developed to address this issue\cite{evans2009direct,dasgupta2002microrheology}.
Here, the expression of the PSD given by Eq.\eqref{eq:32} can be
used to yield rheological parameters of viscoelastic fluids where
our model is suitable (such as a ``Boger'' fluid), and these parameters
can then be employed to get $G^{*}(\omega)$ using \eqref{eq:20}.
Note that this treatment, and the final expression, significantly
resembles the power spectral density method which is widely used to
measure viscosity of Newtonian fluids using optical tweezers. In Fig.\ref{fig:3}
we show that $G^{*}(\omega)$, calculated on one hand using Eq.\eqref{eq:41},
and on the other hand using Eq.\eqref{eq:20} are same. If the parameters
$\lambda,\,\mu_{p},\,\mu_{s}$, and $k$ can be obtained by fitting
Eq.\eqref{eq:32} to the PSD then the method is the same as the power
spectral density approach that is often used to perform microrheology
in viscous fluids. 

\section{Conclusions}

In conclusion, we determine the frequency-dependent viscosity for
a complex fluid from a microscopic origin, where we take into account
the boundary conditions of fluid mechanics including incompressibility
and low Reynold's number as is relevant for microrheology applications.
Thus, we start from the Stokes-Oldryod-B model, which we linearize
assuming small external perturbations. A comprehensive solution yields
an expression for frequency dependent viscosity $\eta(\omega)$, which
we compare to that obtained using the generalized Maxwell (Jeffrey's)
model. We observe that akin to the parameters $\eta_{0}$ and $\eta_{\infty}$
which represent the zero and infinite frequency values of the solvent
in Jeffrey's model, our model yields the parameters $\mu_{p}$ and
$\mu_{s}$, which are the polymer chain and solvent zero-frequency
viscosities, respectively. Interestingly, our model couples the polymer
relaxation time scale $\lambda$ to $\mu_{p}$, which is finally responsible
for the frequency dependence of the viscosity of the complex fluid.
In contrast, the time-scale $\tau_{M}$ appearing in the Jeffrey's
model - which marks the transition from elastic to viscous behaviour
- is coupled with the elastic modulus $G_{\infty}$of the material.
This difference arises from the fact that the Maxwell model does not
consider the contributions of the polymer chains and the solvent separately
- it being based on Maxwell's initial assumption of a viscoelastic
material being made up of a viscous element (represented by $\eta_{0}$)
having an elastic component ($G_{\infty}$). The contribution of the
polymer chains is somewhat ad hoc, with a so-called 'background' viscosity
$\eta_{\infty}$ being added to the frequency dependent term for the
final expression of $\eta(\omega)$. On the other hand, our treatise
makes no such assumption of this form, and instead starts with the
well-known equations for a ``Boger'' elastic fluid, where we have
polymer particles being immersed in a solvent \cite{james2009boger}.
Yet, the final expression we obtain is similar to that obtained using
the Jeffrey's model which justifies the intuition involved in the
latter. After having obtained the expression for $\eta(\omega)$,
we proceed to use it in solving the Langevin equations for Brownian
motion for a free particle, and a particle confined in a harmonic
potential well as is the case in optical tweezers. We determine the
various parameters for particle dynamics including the MSD, and autocorrelations
of the position fluctuations in time and frequency domain. Most importantly,
we observe that the MSD reduces to the well-known expression for a
viscous fluid in the diffusive limit when we have $\mu_{p}\:\rightarrow0$
and $\lambda\:\rightarrow\infty$. This acts as an important consistency
check and gives us confidence in the correctness of our theoretical
treatment. Our expression for $\eta(\omega)$ will prove to be very
useful in microrheology and measurements of viscoelasticity of complex
fluids, which we are presently performing in our laboratory. We would
also like to point out that the entire theoretical treatment has been
performed in the limit of instantaneous hydrodynamic friction, where
we have neglected vorticity diffusion in the fluid which will lead
to retardation effects. This is what we intend to report in the immediate
future, where we will also attempt to solve the entire non-linear
Stokes-Oldroyd-B equations, that should therefore provide a general
form for $\eta(\omega)$.
\begin{acknowledgments}
This work was supported by IISER Kolkata, an autonomous teaching and
research institute supported by the Ministry of Human Resource Development,
Govt. of India, and the Institute of Mathematical Sciences, Chennai,
supported by the Department of Atomic Energy, Govt. of India.
\end{acknowledgments}

\bibliographystyle{apsrev4-1}
\addcontentsline{toc}{section}{\refname}\bibliography{Micro_ref}

\end{document}